# Low-Q whispering gallery modes in anisotropic metamaterial shells

Ana Díaz-Rubio, Jorge Carbonell, Daniel Torrent and José Sánchez-Dehesa*

*Wave Phenomena Group, Department of Electronics Engineering, Universitat Politècnica de València, Camino de Vera, s.n.(Building 7F), ES-46022 Valencia, Spain*

## Abstract

Anisotropic and inhomogeneous metamaterial shells are studied in order to exploit all their resonant mode richness. These multilayer structures are based on a cylindrical distribution of radially dependent constitutive parameters including an inner void cavity. Shell, cavity and whispering gallery modes are characterized, and special attention is paid to the latter ones. The whispering gallery modes are created at the boundary layers of the shell with the background and energy localization is produced with highly radiative characteristics. These low-Q resonant states have frequencies that are independent of the shell thickness. However, their quality factors can be controlled by the number of layers forming the shell, which allows confining electromagnetic waves at the interface layers (internal or external), and make them suitable for the harvesting of electromagnetic energy.

* Corresponding author: jsdehesa@upv.es

# I. Introduction

Textured surfaces are a means of guiding, absorbing, or reflecting electromagnetic waves. Therefore, the so-called surface or interface modes are a powerful alternative to control the propagation, harvesting or redirection of electromagnetic waves[1–3]. These phenomena can be selective in frequency or broadband and give rise to a high number of applications in virtually all spectral bands. To mention just a few recent examples of studies, we can cite frequency selective surfaces for microwaves[4,5], fishnet structures and metasurfaces at terahertz frequencies[6,7], or surface plasmons in optics[8,9]. One of the key underlying drivers in this topic is the use of building blocks based on geometrical arrangements of unitary constituents at a sub-wavelength scale. At the same time, this research context has also been boosted by the rise of metamaterials. These artificial constructs are based on microstructured media (as compared to the operation wavelength) to synthesize a wide variety of effective constitutive parameters, those that govern the behavioral characteristics of electromagnetic waves. With metamaterials it is possible to tailor constitutive parameters (namely permittivity and permeability) so as to achieve negative, null or even extreme values for them. These parameters can in turn be related to other effective material magnitudes like the refractive index or the impedance. Although the 'material' concept usually entails a bulk dimension, opposed to a two dimensional system or surface, both areas share a number of common problems and possibilities.

Linked to the discussion above, this work analyzes the presence of interface or whispering gallery modes in anisotropic metamaterial shells. In particular, we focus on cylindrical shells based on Radial Photonic Crystals (RPC)[10,11]. In comparison with the so called Circular Photonic Crystals (CPC)[12,13], the RPCs are multilayered structures that are invariant under radial translations and verify the Bloch's theorem. The last property is possible in cylindrical coordinates by using anisotropic and radially dependent constitutive parameters, which are feasible using the metamaterial concept. We are interested in finite RPC or RPC-shells consisting of a central cavity and a few layers of RPC. The resonant modes associated to these anisotropic metamaterial shells allow designing devices for different types of applications, like beam shaping shells or position and frequency detectors[11,14]. In addition to the Fabry-Perot like modes located in the RPC shell, cavity modes exist at the central cavity whose features are equivalent to those predicted for cylindrical cavities[15], but now the localization is due to the bandgap created by the RPC-shell. Finally, whispering gallery (WG) modes are the third type of resonant modes existing in these structures and they are localized close to the interfaces at the inner and outer boundaries of the RPC-shell with the background. This main purpose of this work is the study of these resonances not previously described. Out from their main features and among others, we should cite that WG modes have resonant frequencies that are independent of the shell thickness. We may also point out that there are significant differences with respect to localization when compared to Tamm states characterized at the surface of dielectric multilayers[16,17].

The article is organized as follows. After this introduction, Section II briefly reviews the main properties of the photonic band structure of RPCs, for the sake of completeness. Then, Section III presents a detailed discussion on how the properties of a finite RPC can be obtained using the Transfer Matrix Method (TMM), an analytical model that is able to describe all the resonant features associated to the metamaterial shells under study. Section IV analyzes some characteristics of WG modes in order to consider them as building units of devices for energy harvesting. Finally, section V summarizes the main findings of this work.

## II. Radial Photonic Crystals: photonic band structure

RPCs are a type of crystals (i.e., they accomplish the Bloch theorem) that can be briefly described as periodic structures with anisotropic and radially dependent constitutive parameters[10]. This definition equally applies to its acoustic counterpart, the Radial Acoustic Crystals (RAC), since there is a formal equivalence between both problems[18]. For a critical comparison between RPC and RAC the reader is addressed to Refs. 10 and 18.

Here we will focus on electromagnetic waves in cylindrical structures and particularly considering TM$^z$ (E$^z$ polarized) modes. The obtained results can be nevertheless generalized to other arrangements.

Let us consider a RPC in two-dimensions (2D) made of alternating layers of type $a$ (thickness $d_a$) and of type $b$ (thickness $d_b$) whose constitutive parameters are the following:

$$\mu_{ra}(r) = \frac{\hat{\mu}_{ra}}{r} = \frac{0.25d}{r}; \qquad \mu_{rb}(r) = \frac{\hat{\mu}_{rb}}{r} = \frac{0.5d}{r}, \qquad (1)$$

$$\mu_{\theta a}(r) = \hat{\mu}_{\theta a} r = \frac{2r}{d}; \qquad \mu_{\theta b}(r) = \hat{\mu}_{\theta b} r = \frac{r}{d}, \qquad (2)$$

$$\varepsilon_{za}(r) = \frac{\hat{\varepsilon}_{za}}{r} = \frac{d}{1.5r}; \qquad \varepsilon_{zb}(r) = \frac{\hat{\varepsilon}_{zb}}{r} = \frac{d}{r}, \qquad (3)$$

where $r$ is the radial distance from the center of the cylindrical shell and $d$ is the period of the multilayered structure ($d = d_a + d_b$). From these values the components of the refractive index tensor are calculated as follows: the radial component is $n_r = \sqrt{\mu_\theta \varepsilon_z}$ , while the angular component is $n_\theta = \sqrt{\mu_r \varepsilon_z}$ . Figure 1(a) schematically shows a truncated RPC made of 5 periods, and Figs. 1(b)-(c) depict the profiles of the parameters described by Eqs. (1) to (3).

It has been proven[10] that the photonic band structure can be calculated as:

$$\cos(Kd) = \cos(k_{aq}d_a)\cos(k_{bq}d_b) - \frac{1}{2}\left(\frac{\hat{\mu}_{\theta b}}{\hat{\mu}_{\theta a}}\frac{k_{aq}}{k_{bq}} + \frac{\hat{\mu}_{\theta a}}{\hat{\mu}_{\theta b}}\frac{k_{bq}}{k_{aq}}\right)\sin(k_{aq}d_a)\sin(k_{bq}d_b) \qquad (4)$$

The solution for this equation using the parameters described in Eqs. (1)-(3) is displayed in Fig. 2(a), where $q$ denotes the symmetry of the modes in the corresponding band.

Let us remark that each band contains modes with well-defined symmetry. In addition, note that modes with symmetry coefficient $q > 0$ have a cutoff frequency higher than zero. Also, part of the $q = 1$ band (with dipolar symmetry modes) is inserted within the first band gap of mode $q =$ 0 (with monopolar modes). Also, let us stress that by changing the material parameters in Eqs. (1)-(3) is possible to perform a bandgap engineering and design photonic structures that fit our needs. For example, in Fig. 2(a) we observe that modes with dipolar symmetry are in the bandgap of the rest of the modes, which implies that only dipolar modes will be excited by external sources in the frequency region [0.38 - 0.55] (in reduced units). This feature can be extremely useful for designing photonic devices based on finite size RPCs. The modeling of these interesting structures is described below.

## III. Metamaterial shells based on 2D RPCs: analytical model

Let us consider now the case of a finite RPC consisting of N periods and a void central cavity of radius $r_{int}$. Figure 1(a) shows a scheme of a finite RPC consisting of 5 periods with an air cavity being two periods thick ($r_{int} = 2d$). The resulting anisotropic and inhomogeneous metamaterial shells are here studied by means of the Transfer Matrix Method[19] (TMM), which is a suitable procedure for describing their rich resonant behavior.

The $TM^z$ polarized field at an arbitrary point ($r$, $\theta$) in the 2D space can be expressed in cylindrical coordinates as:

$$E_z(r,\theta) = \sum_q E_q(r)e^{iq\theta} \quad , \tag{5}$$

In homogeneous and isotropic media, electric fields can be represented by a linear combination of Bessel and Hankel functions:

$$E_q(r,\theta) = \left[C_{iq}^+ H_q(k_i r) + C_{iq}^- J_q(k_i r)\right]e^{iq\theta} \quad , \tag{6}$$

where $k_i^2 = \omega^2 \varepsilon_i \mu_i$ with $i = 1,2$. Materials 1 and 2 are the materials inside the inner cavity and in the external background, respectively. In turn, electric fields inside the RPC shell ($r > r_i$) can be cast as[10,11,14]:

$$E_q(r,\theta) = \left[(C_q^+)_{ln} e^{ik_{lq}(r'-nd)} + (C_q^-)_{ln} e^{-ik_{lq}(r'-nd)}\right]e^{iq\theta} \quad , \tag{7}$$

where

$$k_{lq}^2 = \hat{\varepsilon}_{zl}\hat{\mu}_{\theta l}\omega^2 - q^2\frac{\hat{\mu}_{\theta l}}{\hat{\mu}_{rl}} \quad,$$

(8)

$r' = r - R_a, \ n = 1,2, \ldots N$ and $l = a, b$.

The electric fields in two consecutive layers are related through the corresponding boundary conditions. Thus, for the E$^z$ polarized modes under study:

$$E_z(r,\theta)|_{r=r_i^+} = E_z(r,\theta)|_{r=r_i^-} \quad,$$

(9)

$$\frac{1}{\mu(r)}\frac{\partial E_z(r,\theta)}{\partial r}\bigg|_{r=r_i^+} = \frac{1}{\mu(r)}\frac{\partial E_z(r,\theta)}{\partial r}\bigg|_{r=r_i^-} \quad.$$

(10)

These conditions are imposed at the interfaces of the unit cell. Therefore, the matrix relating the complex amplitudes of the plane waves in a *b*-layer with those of the equivalent layer of the next unit cell is:

$$\begin{pmatrix}(C_q^+)_{bn-1}\\(C_q^-)_{bn-1}\end{pmatrix} = \begin{pmatrix}A & B\\C & D\end{pmatrix}\begin{pmatrix}(C_q^+)_{bn}\\(C_q^-)_{bn}\end{pmatrix} \quad,$$

(11)

where the transmission matrix *ABCD* elements are:

$$A = e^{-ik_{bq}d_b}\left[\cos(k_{aq}d_a) - \frac{1}{2}i\left(\frac{\hat{\mu}_{\theta b}}{\hat{\mu}_{\theta a}}\frac{k_{aq}}{k_{bq}} + \frac{\hat{\mu}_{\theta a}}{\hat{\mu}_{\theta b}}\frac{k_{bq}}{k_{aq}}\right)\sin(k_{aq}d_a)\right] \quad,$$

(12)

$$B = e^{ik_{bq}d_b}\left[-\frac{1}{2}i\left(\frac{\hat{\mu}_{\theta b}}{\hat{\mu}_{\theta a}}\frac{k_{aq}}{k_{bq}} - \frac{\hat{\mu}_{\theta a}}{\hat{\mu}_{\theta b}}\frac{k_{bq}}{k_{aq}}\right)\sin(k_{aq}d_a)\right] \quad,$$

(13)

$$C = e^{-ik_{bq}d_b}\left[\frac{1}{2}i\left(\frac{\hat{\mu}_{\theta b}}{\hat{\mu}_{\theta a}}\frac{k_{aq}}{k_{bq}} - \frac{\hat{\mu}_{\theta a}}{\hat{\mu}_{\theta b}}\frac{k_{bq}}{k_{aq}}\right)\sin(k_{aq}d_a)\right] \quad,$$

(14)

$$D = e^{ik_{bq}d_b}\left[\cos(k_{aq}d_a) + \frac{1}{2}i\left(\frac{\hat{\mu}_{\theta b}}{\hat{\mu}_{\theta a}}\frac{k_{aq}}{k_{bq}} + \frac{\hat{\mu}_{\theta a}}{\hat{\mu}_{\theta b}}\frac{k_{bq}}{k_{aq}}\right)\sin(k_{aq}d_a)\right] \quad.$$

(15)

For the case of a RPC shell made of *N* unit cells, the following relation applies:

$$\begin{pmatrix}(C_q^+)_{b0}\\(C_q^-)_{b0}\end{pmatrix} = \begin{pmatrix}A & B\\C & D\end{pmatrix}^N\begin{pmatrix}(C_q^+)_{bN}\\(C_q^-)_{bN}\end{pmatrix} \quad.$$

(16)

The continuity conditions at the interface between the inner cavity (medium 1) and the RPC shell produce the following transition matrix:

$$\begin{pmatrix} C_{1q}^+ \\ C_{1q}^- \end{pmatrix} = \frac{i\pi\mu_1}{2} \begin{pmatrix} J_q'(k_1R_a) & -J_q(k_1R_a) \\ -H_q'(k_1R_a) & H_q(k_1R_a) \end{pmatrix} \begin{pmatrix} 1 & 1 \\ Z_1 & -Z_1 \end{pmatrix} \begin{pmatrix} (C_q^+)_{b0} \\ (C_q^-)_{b0} \end{pmatrix} , \quad (17)$$

where $Z_1 = i\frac{\mu_1 k_{bq}}{k_1 \hat{\mu}_{\theta b} R_a}$ .

The boundary conditions at the interface between the RPC shell and the external background (medium 2) give the second transition matrix:

$$\begin{pmatrix} (C_q^+)_{bN} \\ (C_q^-)_{bN} \end{pmatrix} = \frac{1}{2} \begin{pmatrix} 1 & 1 \\ 1 & -1 \end{pmatrix} \begin{pmatrix} H_q(k_2R_b) & J_q(k_2R_b) \\ Z_2H_q'(k_2R_b) & Z_2J_q'(k_2R_b) \end{pmatrix} \begin{pmatrix} C_{2q}^+ \\ C_{2q}^- \end{pmatrix} , \quad (18)$$

with $Z_2 = -i\frac{k_2\hat{\mu}_{\theta b}R_b}{\mu_2 k_{bq}}$ .

Finally, the complex amplitudes of the E-field in the inner cavity and in the external background are related by the overall relation:

$$\begin{pmatrix} c_{1q}^+ \\ c_{1q}^- \end{pmatrix} = \begin{pmatrix} M_{11} & M_{12} \\ M_{21} & M_{22} \end{pmatrix} \begin{pmatrix} c_{2q}^+ \\ c_{2q}^- \end{pmatrix} , \quad (19)$$

$$M = \frac{i\pi\mu_1}{4\sin(Kd)} \begin{pmatrix} J_q'(k_1R_a) & -J_q(k_1R_a) \\ -H_q'(k_1R_a) & H_q(k_1R_a) \end{pmatrix} \begin{pmatrix} A_N & B_N \\ C_N & D_N \end{pmatrix} \begin{pmatrix} H_q(k_2R_b) & J_q(k_2R_b) \\ Z_2H_q'(k_2R_b) & Z_2J_q'(k_2R_b) \end{pmatrix} . \quad (20)$$

The *ABCD* matrix elements are calculated by a method reported previously[20] and can be written as:

$$A_N = \left(2\cos(Kd) + \left(\frac{\hat{\mu}_{\theta b}}{\hat{\mu}_{\theta a}}\frac{k_{aq}}{k_{bq}} - \frac{\hat{\mu}_{\theta a}}{\hat{\mu}_{\theta b}}\frac{k_{bq}}{k_{aq}}\right)\sin(k_{aq}d_a)\sin(k_{bq}d_b)\right)\sin(NKd) - 2\sin((N-1)Kd) , \quad (21)$$

$$B_N = -2i\left(\cos(k_{aq}d_a)\sin(k_{bq}d_b) + \frac{\hat{\mu}_{\theta a}}{\hat{\mu}_{\theta b}}\frac{k_{bq}}{k_{aq}}\sin(k_{aq}d_a)\cos(k_{bq}d_b)\right)\sin(NKd) , \quad (22)$$

$$C_N = -2Z_1i\left(\cos(k_{aq}d_a)\sin(k_{bq}d_b) + \frac{\hat{\mu}_{\theta b}}{\hat{\mu}_{\theta a}}\frac{k_{aq}}{k_{bq}}\sin(k_{aq}d_a)\cos(k_{bq}d_b)\right)\sin(NKd) , \quad (23)$$

$$D_N = Z_1\left(2\cos(Kd) - \left(\frac{\hat{\mu}_{\theta b}}{\hat{\mu}_{\theta a}}\frac{k_{aq}}{k_{bq}} - \frac{\hat{\mu}_{\theta a}}{\hat{\mu}_{\theta b}}\frac{k_{bq}}{k_{aq}}\right)\sin(k_{aq}d_a)\sin(k_{bq}d_b)\right)\sin(NKd) - 2\sin((N-1)Kd) , \quad (24)$$

where $\cos(Kd)$ is the dispersion diagram described by Eq. (4).

The matrix *M* in Eq. (20) is used to determine the transmittance ($T_q$) and the reflectance ($R_q$) of modes with *q*-symmetry from the RPC shell. Their expressions are:

$$T_q = 1/M_{11}, \qquad R_q = M_{21}/M_{11}. \quad (25)$$

The quality factor ($Q$) of a given resonant mode can be also obtained from the matrix element $M_{11}(\omega)$ and involves the calculation of the complex frequencies $\omega_R$ that cancel this matrix element: $M_{11}(\omega_R) = 0$ where $\omega_R = \omega_0 - i\alpha$. Then, the $Q$-factor can be calculated from the real and imaginary parts of the resonance frequency as $Q = \frac{\omega_0}{2\alpha}$.

### A) Application of the TMM to a practical case

The TMM has been applied to the finite RPC shell depicted in Fig. 1(a). The profiles of the constitutive parameters are shown in Figs. 1(b) and 1(c). Note that we employ the same constitutive parameters than that employed in the RPC studied in Sect. II for the sake of comparison. The values of the constitutive parameters (in relative units) within the shell always remain positive and roughly vary between 0.1 and 10.

The transmission properties of the shell are depicted in Fig. 2(b), where the different curves represent the transmittance coefficients, $T_q$, for the total transmitted electrical field ($E_z^t$), whose expression is:

$$E_z^t(r,\theta) = \sum_q A_q^0 T_q H_q(k_0 r) e^{iq\theta} \quad , \qquad (26)$$

where $A_q^0$ represent the amplitudes of the incident wave modes, $T_q$ give information about the interaction between EM waves and the RPC shell, and $H_q$ are the Hankel functions. Note that a given $T_q$ curve is specifically related with the allowed band with the same $q$-symmetry in the dispersion diagram shown in Fig. 2(a). The peaks observed in a selected $T_q$ spectrum represent the resonant modes with $q$-symmetry in the shell. The modes are classified as follows.

### A.1) Fabry-Perot like modes

If the peaks in the coefficients $T_q$ appear at frequencies contained within the photonic bands of the corresponding dispersion relation (see Fig. 2(a)), they are produced by a Fabry-Perot (FP) interference phenomenon due to the shell finite thickness. Figure 3(a) plots, as an example, the E-field pattern of a FP mode with dipolar symmetry ($q$=1). Note that the field is mainly located inside the shell; i.e., within the radius range $r_{int} < r < 5d$. The FP-like resonances have been widely studied in previous works. For a detailed discussion of their properties and their potential application the reader is addressed to previous references[10,11].

*A.2) Cavity modes*

When the peaks in the $T_q$ curves appear within the bandgap of the photonic band with $q$-symmetry, they represent modes that are confined in the central void region. For example, Fig. 2(b) shows that the C1 peak appearing in the profile of the $q = 1$ spectrum (dashed line) corresponds to a cavity mode with dipolar symmetry. C1 has frequency 0.1971 which is below the cutoff of the corresponding band in Fig. 2(a). In a similar manner, peaks C2 and C3 with respective frequencies 0.3157 and 0.4156, are due to the presence of cavity modes with quadrupolar ($q = 2$) and hexapolar symmetry ($q = 3$).

Figure 3(b) displays as an example the E-field pattern corresponding to the cavity mode with quadrupolar symmetry ($q = 2$). Let us stress that cavity modes are strictly related to the size of the inner cavity and, as it is shown in Fig. 3(b), they are strongly localized inside the cavity. Their properties will be further discussed in Sec. IV.

*A.3) Whispering gallery modes*

In addition to the resonant modes previously described we have observed additional features in the transmission spectra that have been associated to whispering gallery (WG) modes. The shoulders annotated in Fig. 2(b) as WG0, WG1, WG2 and WG3 are produced by resonant modes characterized by having their E-field mainly localized in the last layer of the shell, as it is usual for WG modes described in the literature. The frequencies of modes WG0, WG1, WG2 and WG3 have been obtained independently using a method based on finite elements[21] and their values are 0.4724, 0.5607, 0.7667 and 0.9442 (in reduced units). These values are in agreement with the frequencies at which the shoulders appear in the corresponding $T_q$ curves (with $q = 0$ to 3).

The WG modes are characterized by two main properties. On the one hand, their frequencies are always within the bandgap of the photonic bands with the same symmetry. On the other hand, they appear in a truncated RPC with a void cavity at its center. In other words, the structures sustaining WG modes are anisotropic shells having two boundaries with the background. For the case under study here, the external and inner borders are at $r_{ext} = 7d$ and $r_{int} = 2d$, respectively.

Figures 3(c) and 3(d) display, as two typical examples, the E-field patterns of WG-type modes with symmetries $q = 0$ and $q = 3$, respectively; WG0 and WG3 in Fig. 2(b). It is shown that E-fields are mainly localized at the shell outer layers.

Figure 4(a) specifically shows the E-field profile, along the diameter crossing the horizontal axis, for the mode WG0 with frequency 0.4724. A comparison with the profiles of the components of the refractive index tensor, which are shown in Fig. 4(b), indicates that localization takes place in the last period, where the last layer is of $b$-type (with $n_r = 1$). At this point it is interesting to recall that the so called Tamm states were observed at the interface between a multilayered dielectric structure and the background[17]. The Tamm states are strongly localized at the last layers of the multilayers due to the high contrast between the refractive index of the last layers and the background. In contrast, our WG modes, these being a kind of surface states with circular symmetry, are slightly localized and, therefore, highly radiative.

WG modes localized in the inner layer of the shell can be also obtained by simply inverting the sequence of alternating $a$- and $b$-type layers. Figure 5(a) shows the case of a WG mode with monopolar symmetry ($q = 0$) that has been obtained using an inner layer $b$-type and an outer layer *a*-type. The inset shows the 2D field pattern of this mode that resonates at 0.4575 (in reduced units), a value very close to that of the WG0 mode localized in the outer layer of the shell. The high concentration of field observed inside the cavity is due to the leaky nature of this mode in combination with the fact that its wavelength is commensurate with the cavity diameter ($\phi = 2r_{int}$); i.e., $\phi \approx 3\lambda/2$. The radial profile of the E-field shown in Fig. 5(a), in comparison with the radial dependence of components $n_r$ and $n_\theta$ (as shown in Fig. 5(b)), let us to conclude that localization of these types of modes is strongly related with regions with high $n_\theta$.

It has been pointed out that a main feature of WG modes found in RPC shells is that they are strongly radiative. In other words, they have very low *Q*-factors, a property of paramount interest in building devices for energy harvesting. The *Q*-factors of WG modes are specifically studied in the next section in comparison with the *Q*-factors of cavity and FP resonances.

## IV. Energy harvesting with whispering gallery modes

Energy absorption and harvesting is a topic of paramount interest where artificially structured materials have shown potential advantages[3,22-24]. In this context, we consider here a specific configuration of RPC-shells that enhance the energy exchange between their resonant WG modes with the background EM fields.

A metamaterial shell is here designed for operating at frequencies around *f* = 3 GHz (i.e., for wavelengths around $\lambda$ = 100 cm). The shells under study have radial period *d* = 2 mm, ($d_a=d_b=d/2$), the central cavity has radius $r_{int}$ = 0.5 mm ($r_{int} = d/4$), and the constitutive parameters of the alternating layers are:

$$\mu_{ra}(r) = \frac{9.6d}{r}; \; \mu_{rb}(r) = \frac{7.2d}{r}, \qquad (27)$$

$$\mu_{\theta a}(r) = \frac{24r}{d}; \; \mu_{\theta b}(r) = \frac{12r}{d}, \qquad (28)$$

$$\varepsilon_{za}(r) = \frac{8.4d}{r}; \; \varepsilon_{zb}(r) = \frac{49d}{r}, \qquad (29)$$

These parameters are selected to produce WG modes with extremely low $Q$-factors. The profiles for these parameters together with those of the components of the refractive index tensor are described in Fig. 6(a) to Fig. 6(c). The constitutive parameters roughly vary between 1 and 80, a stronger variation than those depicted in Figure 1.

For the sake of comparison, the $Q$-factors of the FP-like modes and cavity modes have been also calculated as a function of the number $N$ of double layers. Results have been obtained using the TMM described in Sec. III and have been compared with the ones calculated using a commercial software[21]. Only radiation losses are considered in all the calculations. The possible dissipative losses associated with the materials have been neglected in this first approach.

Figures 7(a) and 7(b) show the results obtained for the resonance frequencies and $Q$-factors, respectively, of the FP-like modes. Modes with dipolar ($q = 1$), quadrupolar ($q = 2$) and hexapolar ($q = 3$) symmetries are studied. The FP-like mode for a given symmetry corresponds to the lowest frequency mode within the pass band. Note the excellent agreement between the values calculated with the TMM (dashed lines) and the commercial software COMSOL (continuous lines). The main features of these modes are the slight variation of the frequency as a function of the number of layers (upper panel) and also the small variation of the $Q$-factor (less than one order of magnitude).

Figure 8 shows the results corresponding to the cavity modes. Figure 8(a) shows that their frequencies remain constant as a function of the number of layers N and are very high in comparison with that of FP modes. High frequency values are explained because of the small dimensions of the cavity. For this type of modes, only TMM calculations are reported since the commercial software, which is based on the finite elements method, is not efficient in simulating large electrical size objects. Figure 8(b) shows that the corresponding $Q$-factors exponentially increase with N. This can be understood as a consequence of the exponential decaying behavior of the E-field within the photonic bandgap. A similar behavior was observed for the cavity modes confined in RSCs. The reader is addressed to a previous work[20] for a detailed discussion of the properties of cavity modes in anisotropic and inhomogeneous acoustic shells.

Let us point out that the behavior of FP-like modes and cavity modes is completely analogous to the corresponding modes studied in the acoustic counterparts of these photonic structures, the metamaterial shells based on RSC[20]. The energy harvesting carried out by EM or acoustic waves would be more efficient with resonant modes with the lowest $Q$-factors. Then, with this goal in mind, cavity modes in a metamaterial shell made of a small number of layers are preferable. However, WG modes can be obtained with lower $Q$-factors as it is explained below.

Figure 9 reports the properties of WG modes localized at the outer shell surface as a function of N. Figure 9(a) shows that their frequencies have a negligible dependence with the shell thickness. Their values for N = 2 (4 layers) are 3.35 GHz (WG1), 4.29 GHz (WG2) and 5.36 GHz (WG3). For the case N = 1 (a double layer) no surface mode was obtained since the bandgap concept cannot apply to a single period shell. Again, results obtained with our analytical TMM are well supported by the numerical experiments using COMSOL.

A very remarkable property of these modes is that their frequencies, as in the case of cavity modes, do not vary with the shell thickness. This would seem counter-intuitive since the frequency of WG modes in cylindrical cavities made of isotropic dielectric medium depends on the optical path determined by the cavity perimeter; that is $\ell = n \times 2\pi R$, where $n$ is the refractive index and $R$ the cavity radius. For the anisotropic and inhomogeneous shell under consideration, Fig. 6(c) shows that $n_r$ is constant within each layer of types $a$ or $b$, their values alternate in consecutive layers between $n_{ra} = 15$ and $n_{rb} = 25$. Since the external layer (of $b$-type) has a higher angular refractive index, compared to the background or the closest shell layer (of *a*-type) it may guide some energy in the angular direction. Conditions exist for these modes to be propagative within the external *b*-layer, and this time again resonant modes appear due to the finite size of the perimeter of this external layer. Refractive index in this perimeter varies as $n_\theta = {}^{ct}/_r$, which is an inverse function of the radial distance, and with *ct* being a constant value depending on the *a*- or *b*-type layer. Now, this refractive index with inverse radial dependence makes that the 'optical path' ($\sim n_\theta \times 2\pi r = ct$) of the outer layer is independent of the size (or number of layers) of the shell. This explains the constant resonant frequencies of the whispering gallery modes and their independence with respect to the number of layers.

Figure 9(b) indicates that the *Q*-factors decrease with the thickness of the shell, an interesting feature that can be useful for energy harvesting. Note that mode WG1 has the lowest *Q*-factor among all the modes here analyzed. For the higher order WG modes, their *Q*-factors are 2 orders of magnitude lower than those calculated for the FP like modes with similar order. In comparison with the *Q*-factors obtained for cavity modes C1 and C2, WG1 has a *Q*-factor comparable with that of C1 at N = 2 while WG2 has a *Q*-factor two orders of magnitude larger (at N = 2). However, for increasing shell thickness WG modes strongly decrease their *Q*-factor up to values several orders of magnitude lower than that of the cavity modes. It can be concluded that WG modes associated with a metamaterial shell thick enough are the best candidates to guarantee an efficient energy exchange with the EM waves to/from the background. It can be said that increasing the radius implies a longer evanescent path of the wave towards the inner layers of the shell (along the radial direction), and effectively a stronger reflection of the wave towards the outer background of the shell. This characteristic effectively increases radiation and exchange with the background from the surface mode in the external layer.

Finally, let us discuss the case of two degenerate WG modes, one being located at the outer boundary and the other at the inner boundary. This case is achieved considering a shell with an odd number of layers (7 layers equivalent to 3 periods and 1 additional *a*-type layer); the total diameter being equal to $\phi = 15$ mm. Figure 10 reports how a point source located at a distance $r_{source} = 20$ mm ($\equiv 10d$) from the shell center illuminates the shell for three slightly different frequencies. It is shown how the WG2 mode at the outer layer is excited at 4.277 GHz while the WG2 mode at the inner layer is excited at 4.323 GHz. Moreover, Fig. 10(b) shows that both modes are simultaneously excited at 4.303 GHz. The frequency differences are much smaller than those depicted in Figs. 4 and 5, since the inner void cavity size is much (electrically) smaller and there is a lower perturbation due to it. Especially interesting is the case depicted in Fig. 10(c), where the resonant mode is linked to the inner interface of the shell with the cavity. This is a representative example of how the WG modes can be used as a mechanism for dramatically improving the energy harvesting from external EM waves and collecting it at the central active area.

## V. Conclusion

We have analyzed the existence of low-$Q$ whispering gallery modes at the interface layers of anisotropic and inhomogeneous metamaterial shells. Conditions for these modes to exist are basically the existence of prohibited band gaps in the dispersion diagram of the shell and interface layers with high refractive index as compared to the background. These modes present the particular feature of having resonant frequencies independent of the shell thickness, a property allowing adapting the quality factor of the resonant modes in a very simple manner. At the same time, energy exchanges with electromagnetic waves in the surrounding background are favored, allowing the use of the anisotropic metamaterial shells as a means of harvesting electromagnetic energy. Particularly, broadband operation by the reported low-$Q$ whispering gallery modes can be obtained by changing the properties of the cavity medium, like its radius and dielectric constant.

## Acknowledgment

The authors acknowledge the financial support of Spanish Ministry MINECO (grant numbers TEC 2010-19751 and Consolider CSD2008-00066), and the U.S. Office of Naval Research (USA).

# References

[1] C. R. Williams, S. R. Andrews, S. A. Maier, A. I. Fernandez-Dominguez, L. Martin-Moreno, and F. J. Garcia-Vidal, Nature Photonics **2**, 175 (2008).

[2] C. M. Watts, X. Liu, and W. J. Padilla, Adv Mater **24**, OP98 (2012).

[3] P. Spinelli, M. A. Verschuuren, and A. Polman, Nature Communications **3**, 692 (2012).

[4] B. A. Munk, *Frequency selective surfaces: theory and design* (John Wiley & Sons, Inc., Hoboken, NJ, USA, 2000), p. 410.

[5] J. Carbonell, E. Lheurette, and D. Lippens, Prog. Electromagn. Res. **112**, 215 (2011).

[6] S. Wang, F. Garet, K. Blary, C. Croenne, E. Lheurette, J. Coutaz, and D. Lippens, J. Appl. Phys. **107**, 074510 (2010).

[7] M. Aznabet, M. Navarro-Cia, S. A. Kuznetsov, A. V. Gelfand, N. I. Fedorinina, Y. G. Goncharov, M. Beruete, O. El Mrabet, and M. Sorolla, Optics Express **16**, 18312 (2008).

[8] W. Barnes, A. Dereux, and T. Ebbesen, Nature **424**, 824 (2003).

[9] J. Pendry, L. Martin-Moreno, and F. Garcia-Vidal, Science **305**, 847 (2004).

[10] D. Torrent and J. Sanchez-Dehesa, Phys. Rev. Lett. **103**, 064301 (2009).

[11] J. Carbonell, A. Diaz-Rubio, D. Torrent, F. Cervera, M. A. Kirleis, A. Pique, and J. Sanchez-Dehesa, Scientific Reports **2**, 558 (2012).

[12] N. Horiuchi, Y. Segawa, T. Nozokido, K. Mizuno, and H. Miyazaki, Opt. Lett. **30**, 973 (2005).

[13] P. Lee, T. Lu, J. Fan, and F. Tsai, Appl. Phys. Lett. **90**, 151125 (2007).

[14] J. Carbonell, D. Torrent, and J. Sánchez-Dehesa, IEEE Transactions on Antennas and Propagation **61**, 755 (2013).

[15] R. F. Harrington, *Time-Harmonic Electromagnetic Fields* (IEEE Press - Wiley Interscience, New York, NY, 1961).

[16] A. P. Vinogradov, A. V. Dorofeenko, S. G. Erokhin, M. Inoue, A. A. Lisyansky, A. M. Merzlikin, and A. B. Granovsky, Physical Review B **74**, 045128 (2006).

[17] T. Goto, A. V. Baryshev, M. Inoue, A. V. Dorofeenko, A. M. Merzlikin, A. P. Vinogradov, A. A. Lisyansky, and A. B. Granovsky, Physical Review B **79**, 125103 (2009).

[18] J. Carbonell, D. Torrent, A. Diaz-Rubio, and J. Sanchez-Dehesa, New J. Phys. **13**, 103034 (2011).

[19] A. Yariv and P. Yeh, *Optical waves in crystals* (Wiley New York, 1984), 5.

[20] D. Torrent and J. Sanchez-Dehesa, New Journal of Physics **12**, 073034 (2010).

[21] Comsol AB (Sweden), Comsol Multiphysics (v. 4.2a), www.comsol.com, 2011.

[22] Y. Yao, J. Yao, V. K. Narasimhan, Z. Ruan, C. Xie, S. Fan, and Y. Cui, Nature Communications **3**, 664 (2012).

[23] C. Navau, J. Prat-Camps, and A. Sanchez, Phys. Rev. Lett. **109**, 263903 (2012).

[24] J. Carbonell, F. Cervera, J. Sanchez-Dehesa, J. Arriaga, L. Gumen, and A. Krokhin, Appl. Phys. Lett. **97**, 231122 (2010).

## Captions

**FIG.1.** (a) Schematic picture of a truncated radial photonic crystal. The resulting metamaterial shell has an inner void cavity of radius $r_{int} = 2d$ and 5 periods of alternating type-$a$ and type-$b$ materials. (b) Profiles of the radial ($\mu_r$) and angular ($\mu_\theta$) permeabilities as a function of the radial distance. (c) Profile of the dielectric permittivity $\varepsilon_z$.

**FIG. 2.** (a) Photonic band structure for the radial photonic crystal described in section II. (b) Calculated transmission coefficients $T_q$ for the 5 period metamaterial shell described in Fig. 1. The peaks in a given $T_q$ curve represent resonant modes with $q$-symmetry. The cavity (C) and whispering gallery (WG) modes are marked with arrows. The unmarked peaks correspond to Fabry-Perot (FP) modes and are mostly located within the shell.

**FIG. 3.** Resonant modes found in the 5 period RPC-shell described in Fig. 1: (a) Fabry-Perot like mode with symmetry $q = 1$ and frequency 0.3101 (in reduced units), (b) cavity mode with symmetry $q = 2$ and frequency 0.3157 (C2 in Fig. 2(b)), (c) whispering gallery mode with symmetry $q = 0$ and frequency 0.4724 (WG0 in Fig. 2(b)), and (d) whispering gallery mode with symmetry $q = 3$ and frequency 0.9442 (WG3 in Fig. 2(b)).

**FIG. 4.** (a) The E-field profile along a diameter section of a WG mode located at the outer layer of the 5 period RPC shell described in Fig. 1. The monopolar mode WG0 with frequency 0.4724 is depicted. Note how the field amplitude exponentially decreases with the separation from the external boundary. The inset shows the E-field pattern in 2D for comparison purposes. (b) Radial dependence of the radial and angular components of the refractive index tensor.

**FIG. 5.** (a) The E-field profile along a diameter section of a WG mode located at the inner layer of the 5 period RPC shell with inverted sequence ($a$-type and $b$-type layers have been exchanged

with respect to Fig. 1). The monopolar WG0 mode with frequency 0.4575 is depicted. Note the high concentration of the field in the cavity due to the leaky nature of this mode. The insets show the E-field pattern in 2D for comparison purposes. (b) Radial dependence of the radial and angular components of the refractive index tensor.

**FIG. 6.** Profiles of the constitutive parameters defining an anisotropic metamaterial shell made of 5 periods of alternating materials and having a central cavity with radius $r_{int} = 2d$. (a) radial and angular permeabilities, $\mu_r$ and $\mu_\theta$, (b) permittivity $\varepsilon_z$, and (c) radial and angular components of the refractive index; $n_r$ and $n_\theta$, respectively.

**FIG. 7.** Frequency variation (a) and quality factor $Q$ variation (b) as a function of the number $N$ of periods for the Fabry-Pérot resonances located in the shell described in Fig. 6. The radius of the cavity being $r_{int} = 0.5$ mm. Results from a commercial software (COMSOL) are compared with the transfer matrix method (TMM).

**FIG. 8.** Frequency variation (a) and quality factor $Q$ variation (b) as a function of the number $N$ of periods for the modes located in the inner void cavity of the shell described in Fig. 6. The radius of the cavity being $r_{int} = 0.5$ mm. Results are obtained with the transfer matrix method (TMM).

**FIG. 9.** Frequency variation (a) and quality factor $Q$ variation (b) as a function of the number $N$ of periods for the whispering gallery resonances located in the shell described in Fig. 6. The radius of the cavity being $r_{int} = 0.5$ mm. Results from a commercial software (COMSOL) are compared with the transfer matrix method (TMM).

**FIG. 10.** E-field patterns of an external point source illuminating an RPC shell and exciting a whispering gallery mode, (a) an external whispering gallery mode at 4.277 GHz, (b) simultaneous excitation of internal and external whispering gallery modes at 4.303 GHz and (c) internal

whispering gallery mode at 4.323 GHz. Here the total number of layers of the shell is 7, with inner and outer layers of the same $a$-type.

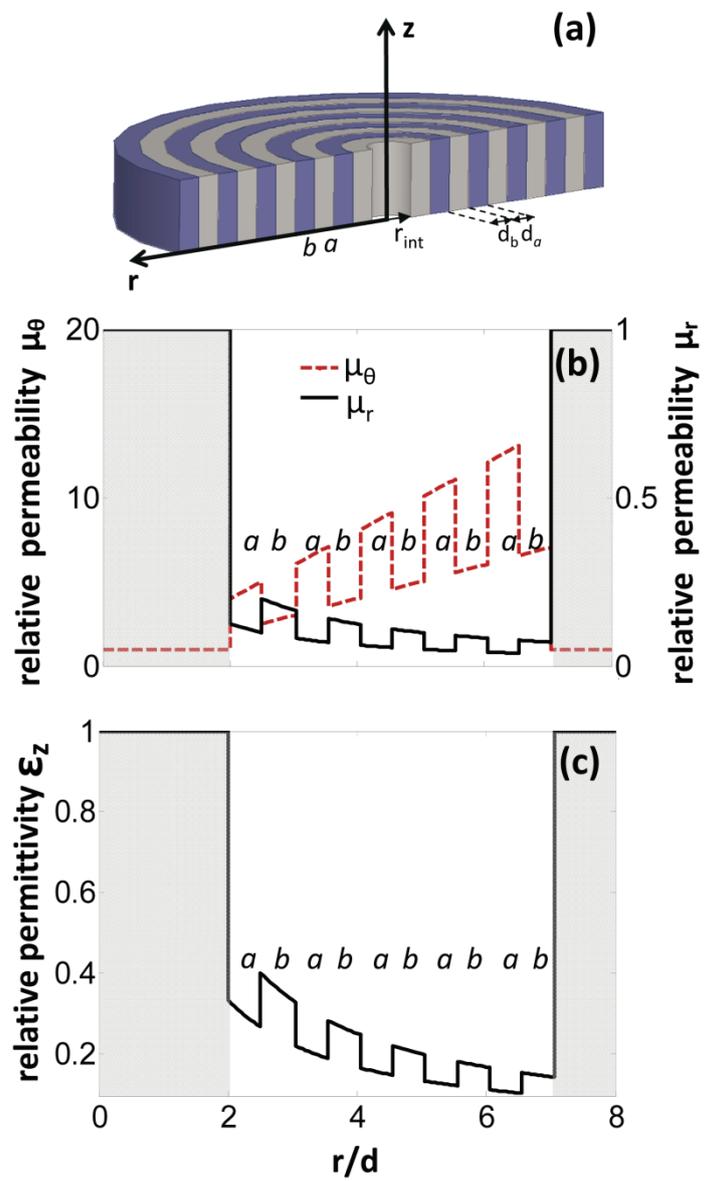

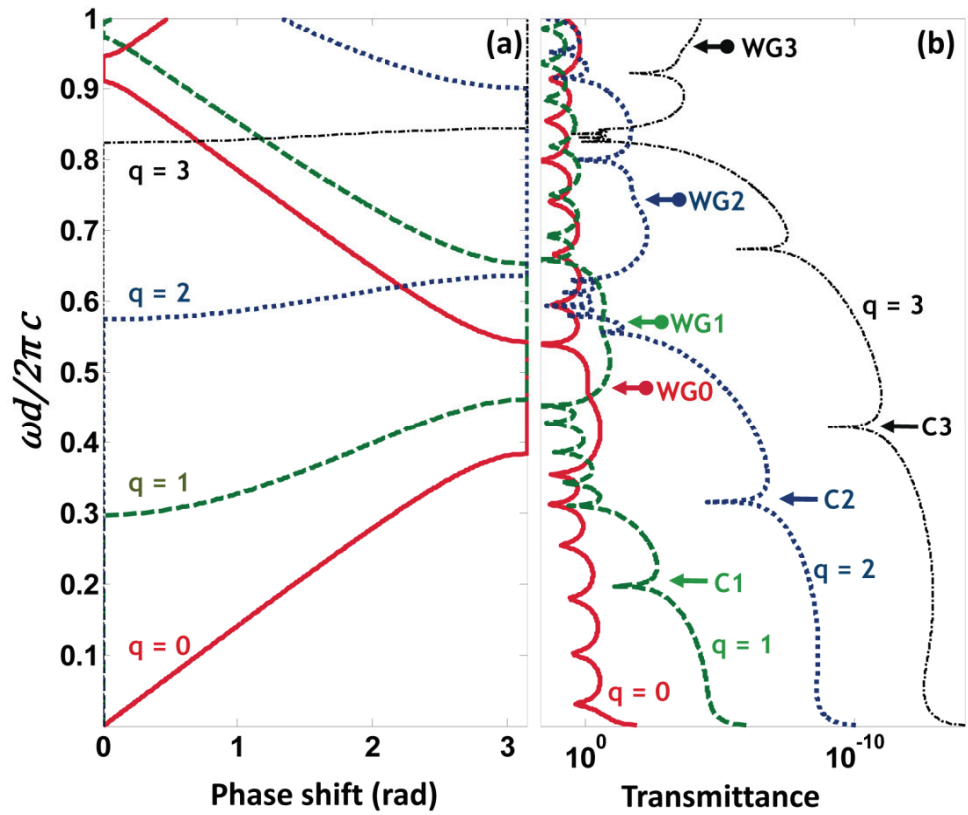

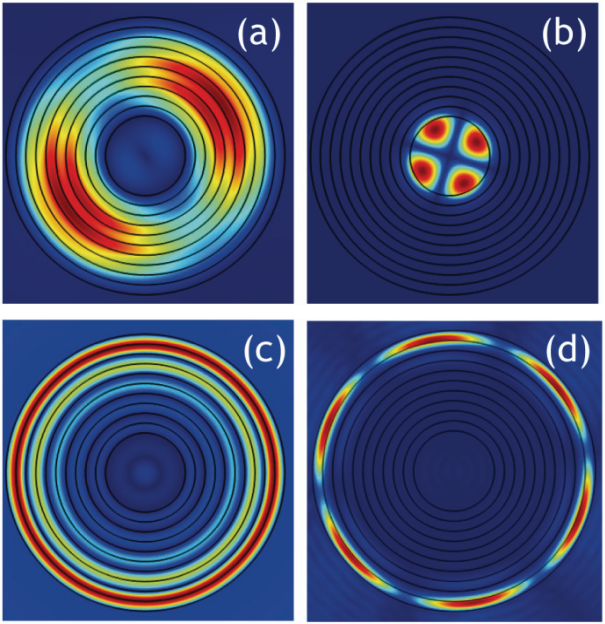

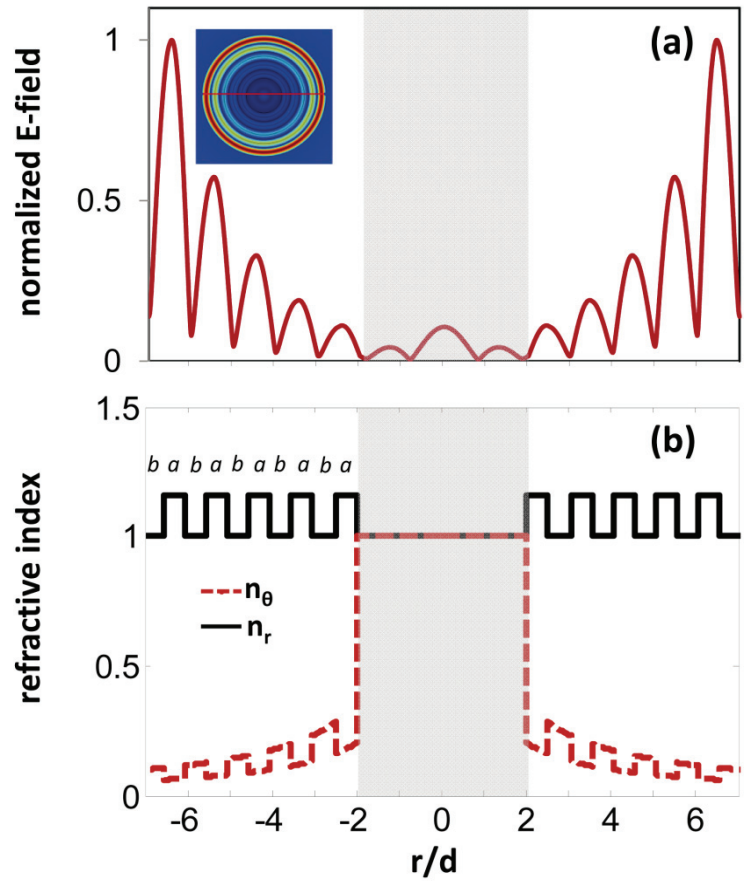

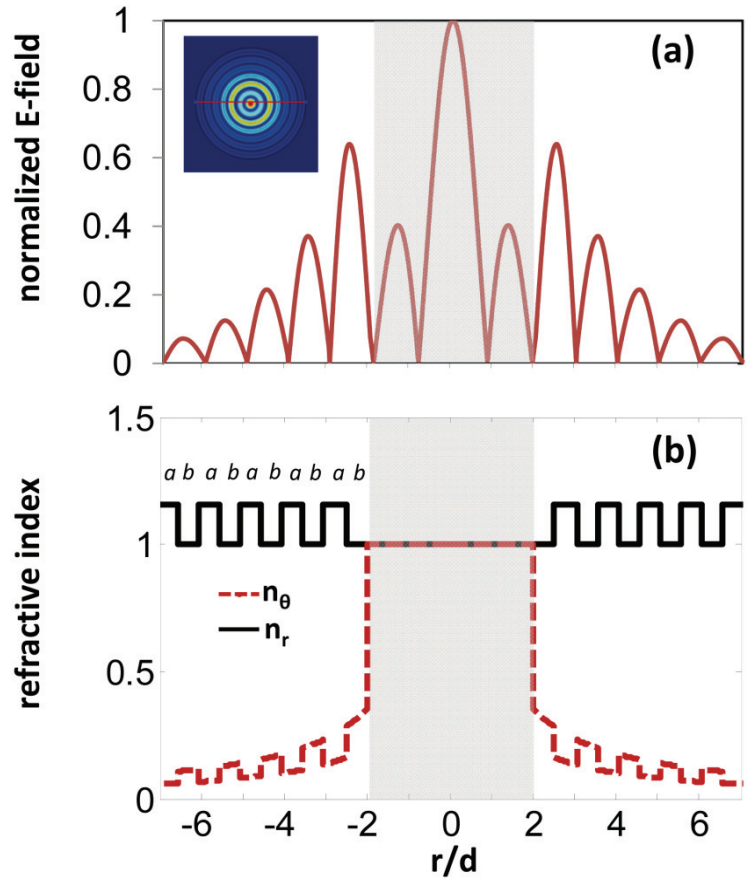

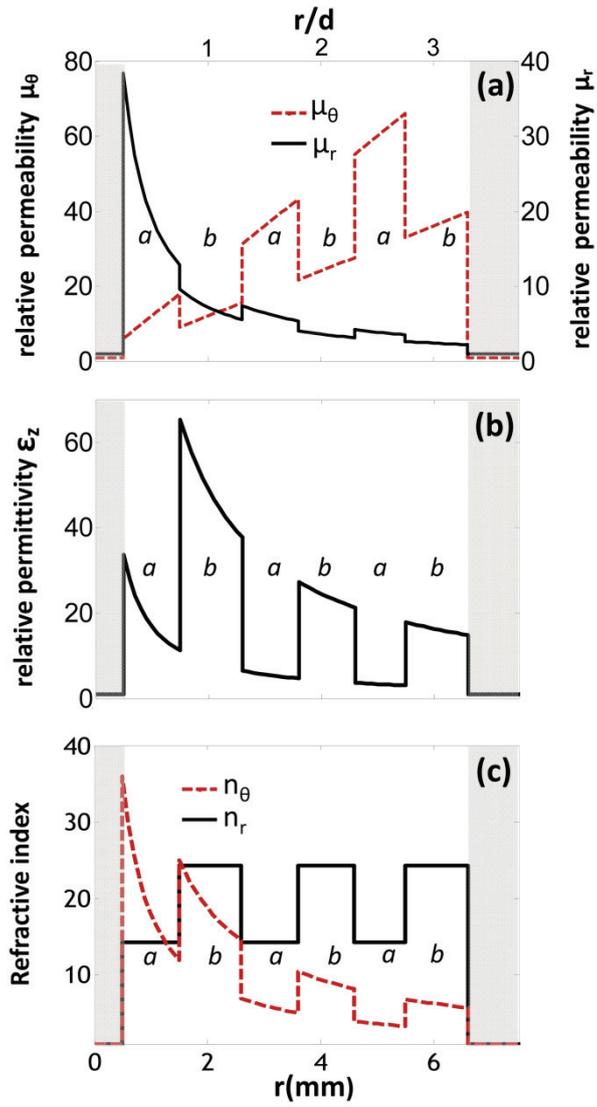

FIGURE 7 (Ana Díaz-Rubio *et al.*, PHYSICIAL REVIEW B)

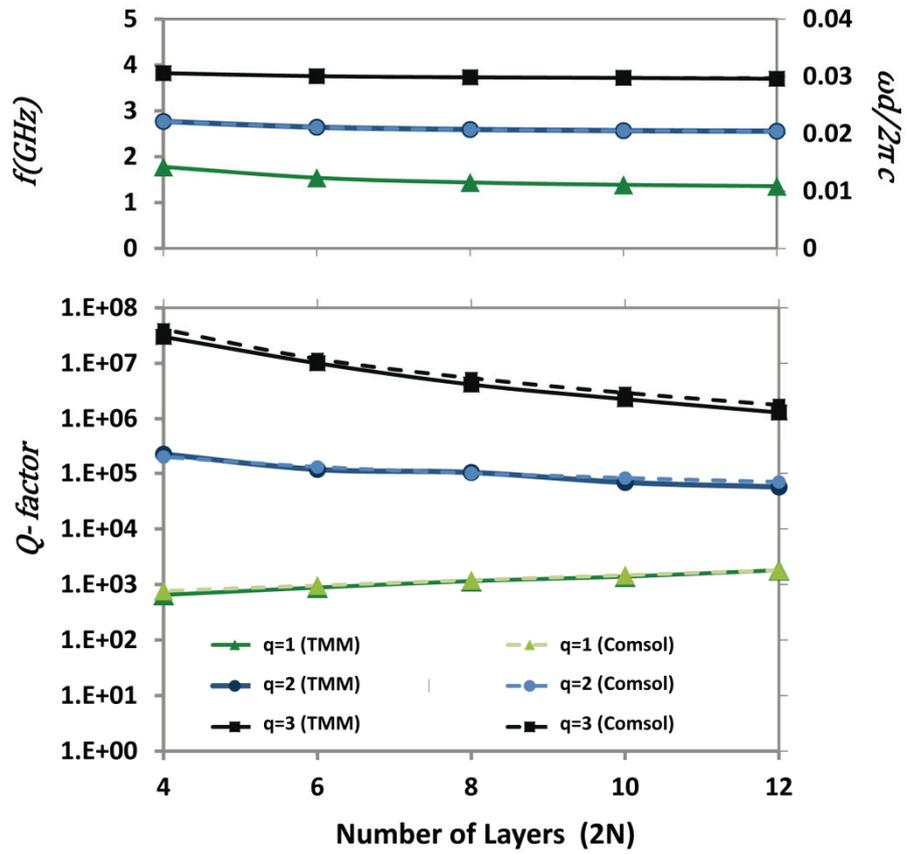

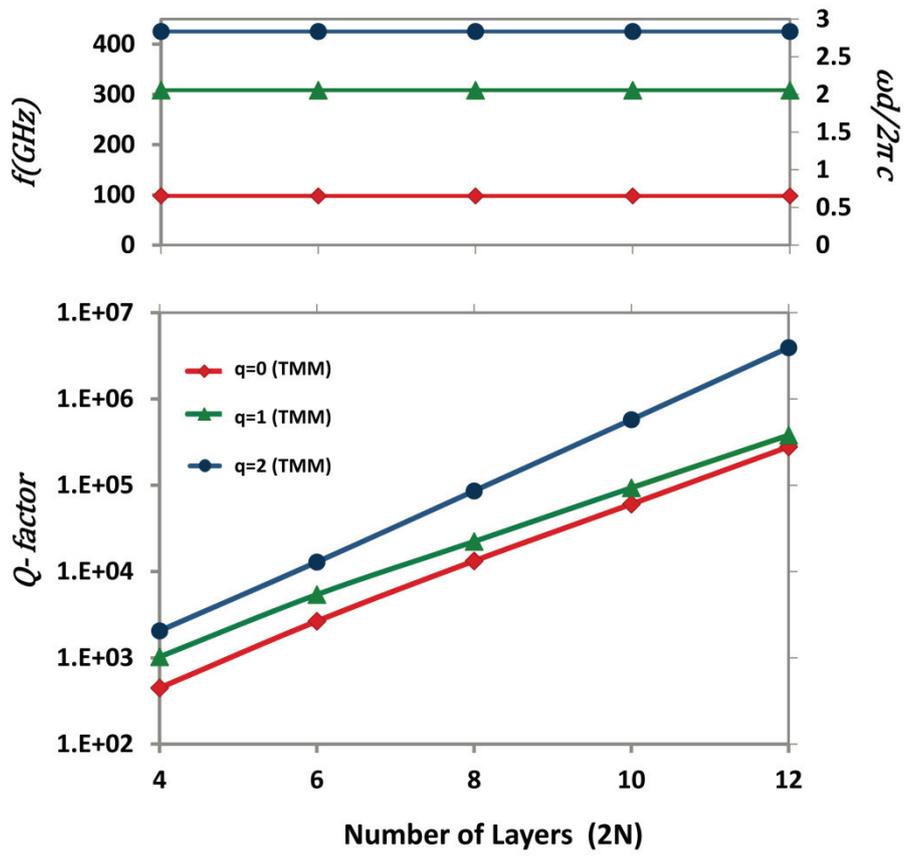

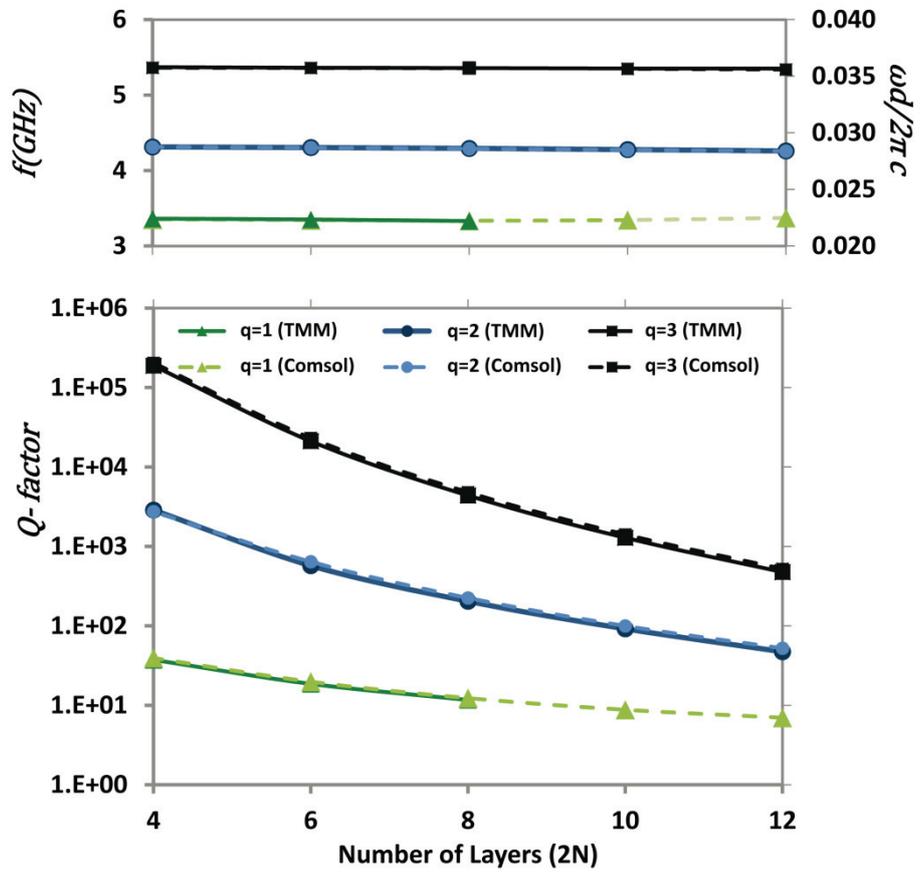

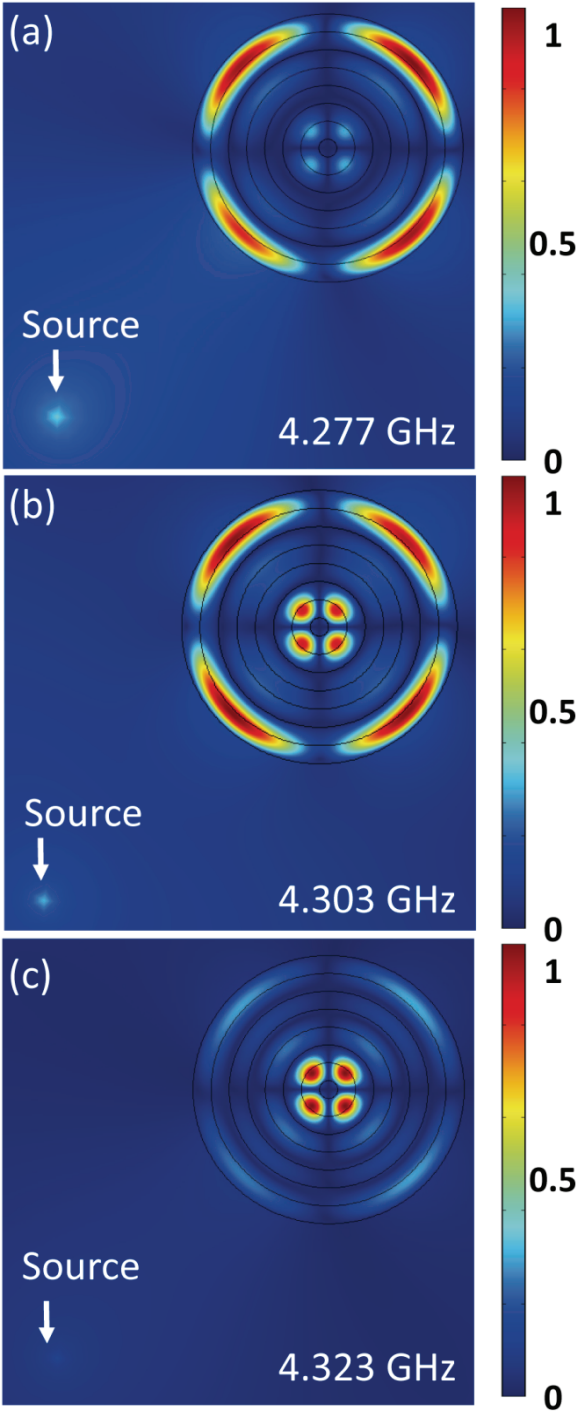